\documentstyle[aaspp4]{article}

\begin{document}     
\title{ Transient Emission From Dissipative Fronts in Magnetized, Relativistic
Outflows. I. Gamma-Ray Flares}

\author{Amir Levinson}
\affil{School of Physics and Astronomy, Tel Aviv University, 
Tel Aviv 69978, Israel}

\begin{abstract}
The transient emission produced behind internal shocks that are 
driven by overtaking
collisions of a magnetized, relativistic outflow is considered.
A self-consistent model capable of describing the structure and 
dynamics of the shocks
and the time evolution of the pair-and gamma-ray distribution functions
is developed and applied to gamma-ray flares in blazars, in the case 
in which gamma-ray production is dominated by inverse Compton scattering
of external radiation (ERC).  The dependence of the flare properties
on magnetic field dissipation rate, intensity of ambient 
radiation, and the thickness of expelled fluid slabs is analyzed.  
It is shown that i) the type of gamma-ray flare produced by the model 
is determined by the ratio of the thickness of ejected fluid slab and 
the gradient length scale of ambient radiation intensity,
ii) the radiative efficiency depends sensitively on 
the opacity contributed by the background radiation, owing to a
radiative feedback, and is typically very high for parameters
characteristic to the powerful blazars, and
iii)  the emitted flux is strongly suppressed at energies 
for which the pair-production optical depth is initially larger
than unity; the time lag and flare duration in this energy range 
increase with increasing gamma-ray energy.  At lower energies, 
flaring at different gamma-ray bands occurs roughly simultaneously,  
but with possibly different amplitudes.  Some observational consequences
are discussed.
										      
\end{abstract}

\section{Introduction}
The discovery of strong, variable gamma-ray sources identified with 
blazars by CGRO has motivated many theoretical investigations concerning 
the emission from relativistic jets, and several models of gamma-ray blazars 
have recently been developed (e.g., Burns \& Lovelace 1982; 
Dermer \& Schlickeiser 1993; Bloom \& 
Marscher 1993; Manneheim 1993; Sikora, et al. 1994; Blandford \& Levinson
1995 [BL95]; Ghisellini \& Madau 1996).  However, most of these 
efforts have been 
devoted to examine spectral properties of blazars using steady-state 
models, which are most suitable for exploring quiescent states.
Although some progress in identifying the emission mechanisms 
in individual sources (e.g., Sambruna et al. 1997; Sikora et al. 1997), 
and constraining the structure and dynamics of relativistic jets on small 
scales (Levinson 1996) has been made, it 
has became widely recognized that quantitative analysis of the growing
body of variability data is crucial for advancing our understanding 
of the nature of relativistic jets further.
Recent observational efforts (e.g., Reich et al. 1993; Maraschi 
et al. 1994; Wagner 1996; Buckley et al. 1996; Takahashi et al. 1996
; Mattox et al. 1997; Wehrle et al. 1997; for a recent review 
see Ulrich 1997) to characterize the transient 
emission in blazars motivate detailed analysis of various variability
mechanisms, and the development of time dependent models.  

Different processes may lead to time variability of blazar emission, 
including sudden changes in particle injection rate and/or magnetic 
field, changes in the bulk speed, and temporal variations of the 
intensity of background radiation in ERC models.  This paper 
investigates the possibility that gamma-ray flares observed
in blazars are produced by internal shocks propagating in a 
magnetized, relativistic jet.  
The work presented in this paper extends
an earlier work by Romanova \& Lovelace (1997; hereafter RL97).  An outline 
of dissipative fronts is given in \S 2.  In \S 3 we present the model and
derive the basic equations.  The results are described in \S 4.  
We conclude in \S 5.
											
\section{Dissipative fronts in magnetized, relativistic outflows}

Temporal fluctuations in the parameters of a MHD outflow lead to 
excitation of waves that steepen into shocks at some distance from the 
fluid ejection point, and the ultimate formation of a front consisting of a 
pair of shocks and a contact discontinuity across which the total 
pressure (kinetic plus magnetic) is conserved.  
The front propagates at a speed intermediate 
between that of the colliding streams and expands at a rate proportional
to the relative velocity between the two shocks.  Consequently, there is
a net energy flow into the front which is balanced
by either adiabatic cooling, owing to the front expansion, or radiative 
cooling, depending upon the conditions in the front.  
In the case of Poynting flux dominated 
outflows, rapid magnetic filed dissipation is required in order for a 
significant fraction of the outflow energy to be converted into kinetic
energy behind the shocks.  Efficient magnetic field dissipation may 
be achieved in strongly turbulent MHD flows (Thompson 1997) 
(an {\it ordered} magnetic field component and an associated 
Poynting flux can still be defined), or in
the case of a sudden reversal of magnetic field lines (Romanova \& 
Lovelace 1992).  

We envision that the Lorentz factor of a fluid expelled
from the central engine increases suddenly from $\Gamma_{+}$ to
$\Gamma_{-}>\Gamma_{+}$, and denote by $\tau_{acc}$ the characteristic 
time change of the outflow parameters (of order the dynamical time in 
the injection region).
The ejection of the fast fluid is assumed to persist for time 
$\tau_{inj}$, after which it abruptly terminates, and the outflow 
Lorentz factor drops to lower values.  As described 
above, this would lead to the creation of forward and reverse shocks
that will propagate across the slow and fast fluid slabs, respectively. 
(The drop in Lorentz factor after time $\tau_{inj}$ will lead also 
to the formation of a rare-faction wave behind the rapid slab.)  
The position $r_o$ at which the front is created has been calculated by 
Levinson \& van Putten (1997) using simple-wave analysis.  As they 
have shown, in the limit
in which the disturbance speed with respect to the rest frame of the 
boundary at which the fluid is injected (henceforth referred to as
injection frame; this might be e.g., the rest frame of the central 
engine ) is highly relativistic, $r_o=\kappa c\tau_{acc}
\Gamma_{+}^2\Gamma_A^2$, where $\Gamma_A$ is the Lorentz factor 
associated with the Alf\'ven speed with 
respect to the fluid rest frame, and $\kappa$ is a numerical factor that
depends on the details of fluid injection ($\kappa=1/3$
when the 4-velocity of the expelled fluid increases according to  
$u=\sinh[\lambda+t/\tau]$; $t\ge0$).  In 
the case of extragalactic jets, 
$\Gamma_{+}$ is typically of order a few.  $\Gamma_A$ may range from 
$\sim$1, in the case of a weakly magnetized flow, to $\ge 10$
beneath the annihilation radius in 
Poynting flux jets (BL95; Levinson 1996).  Thus, if we associate 
$c\tau_{acc}$ with the gravitational radius of the putative black 
hole, we anticipate $r_o$ to lie in the range between $10^{15}$ and 
$10^{18}$ cm in blazars.

The creation of fronts in non-relativistic, hydrodynamic flows has been 
considered by Raga et al. (1990) in the context of HH objects.  The basic
idea has later been generalized and applied to Poynting flux 
jets in blazars by RL97.  These authors proposed a scenario in which the 
gamma-ray and synchrotron outbursts 
often seen in blazars are ascribed to cooling of relativistic 
electrons accelerated inside highly dissipative fronts 
propagating in a magnetically dominated jet.  The physical processes
included in their model are inverse Compton scattering of external
radiation, synchrotron and SSC emission.  In their treatment, however, the 
spectrum of emitting particles is assumed a priori, and is characterized
by two break energies which are allowed to evolve with time.  
The structure of the front is not calculated self-consistently, but rather
the front is assumed to expand freely with the speed of sound.  As we show,
the front structure can be significantly altered by radiative losses.
Moreover, their model does not account for absorption of gamma-rays 
escaping the front by pair production on ambient photons ahead of the 
front.  As shown below, this process may considerably affect the emitted 
spectrum, particularly in the powerful blazars.  

In this paper we generalize RL97 treatment in several important ways.  Firstly,
the time evolution of the energy distribution of pairs and 
gamma-rays in the front is calculated self-consistently, by numerically 
integrating the coupled kinetic and MHD equations.  Secondly, 
the effect of radiative losses on the evolution of the front structure is 
taken into account.  Thirdly, pair cascades inside and ahead of the 
front are accounted for properly.  Fourthly, finite length outbursts are
considered.  And finally, we use different initial conditions.   
Our model involves the following key parameters: i) the 
dissipation rate of magnetic energy, ii) the maximum injection energy of
electrons, denoted by $E_{emax}$, iii) the fraction $\eta$ of dissipation
energy that is tapped to the injection of electrons to $E_{emax}$, and iv)
the ratio of the thickness of ejected fluid slab and the gradient length
scale of the intensity of background radiation.  We suppose that on 
scales of interest
the outflow material is dominated by e$^{\pm}$ plasma, and ignore possible
effects associated with entrainment of ions (which are likely to be present
in the MHD wind confining the jet) on the front dynamics.  
In the present work, we consider only 
gamma-ray flares produced by the ERC mechanism.  The extension 
of this work to encompass synchrotron and SSC flares will be presented
in a follow-up paper.		

\section{The model and basic equations}
\subsection{Front dynamics}

The stress-energy tensor of a magnetized flow can be written as a 
sum of two contributions: the energy-momentum tensor associated 
with the electromagnetic field, $T^{\mu\nu}_{em}$, and 
the energy-momentum tensor associated with the system of particles,
$T^{\mu\nu}_{s}$. $T^{\mu\nu}_{s}$ and the particle flux, $N^{\mu}$,
are related to the electron distribution function,
defined in \S 4.2 below through,
\begin{equation}
N^{\mu}=nu^{\mu}=\int{p^{\mu}f_s(p,x)\frac{d^3p}{p^o}},\\
\label{eq: Nu}
\end{equation}
and
\begin{equation}
T^{\mu\nu}_{s}=\int{p^{\mu}p^{\nu}f_s(p,x)\frac{d^3p}{p^o}},
\label{eq: Tu}
\end{equation}
where $n$ is the proper density and $u^{\alpha}$ the 
fluid 4-velocity.
In terms of the proper gas pressure, $p$, 
specific enthalpy, $h$, magnetic induction, 
$b^{\alpha}=\tilde{F}^{\alpha\beta}u_{\beta}$, with 
$\tilde{F}^{\alpha\beta}$ being the dual electromagnetic 
tensor, and rest mass density $\rho=m_en$, the stress-energy tensor 
takes the form
\begin{equation}
T^{\mu\nu}=T^{\mu\nu}_{s}+T^{\mu\nu}_{em}=\rho h^{\ast} u^{\mu}u^{\nu}
+p^{\ast}g^{\mu\nu}-b^{\mu}b^{\nu}.
\end{equation}
Here $g_{\alpha\beta}$ denotes the metric tensor, $h^{\ast}=h+b^2/(\rho 
c^2)$, $p^{*}=p+b^2/2$, and $b^2=b^{\alpha}b_{\alpha}$. 

Let $Q$ and $S^{\nu}$ denote the rate of change of the electron density due to 
e$^{\pm}$ pair creation and annihilation, and the rate of change 
of the $\nu$ component of the stress-energy tensor resulting from 
radiative losses, respectively, and $Q_b^{\beta}$ denotes the source term 
associated with magnetic field dissipation.  The fluid equations can 
then be expressed as,   

\begin{eqnarray}
\partial_{\alpha}T^{\alpha\beta}=S^{\nu},\nonumber \\
\label{eq: MHD}
\partial_{\alpha}N^{\alpha}=Q,\\
\partial_{\alpha}(u^{[\alpha}b^{\beta]})=Q_{b}^{\beta}.\nonumber 
\end{eqnarray}
The terms $Q$ and $S^{\mu}$ are expressed in \S 4.3 below in terms of 
the Boltzmann collision operators.  The evaluation of these source terms 
at any given time involves the integration of the kinetic equations 
(eqs. [\ref{eq: Bolz-e}] and [\ref{eq: Bolz-g}]) for the pairs and photons.
											
We consider the evolution of a front produced by a collision of fast 
and slow fluids having velocities $\beta_{-}$ and $\beta_{+}$, 
and corresponding Lorentz factors $\Gamma_{-}$ and $\Gamma_{+}$, 
respectively.  We suppose that each shock decays after it 
crosses the corresponding fluid slab, at which point energy 
deposition into that slab ceases.  At times shorter than 
the shock crossing times, the set of equations (\ref{eq: MHD}) 
can be integrated over a portion 
of space-time containing the front, using the generalized Gauss's theorem, 
to obtain a set of ordinary differential equations governing the time 
evolution of the front quantities.  The derivation of the front 
equations is given in the appendix.  In what follows,  
subscript minus (plus) refers to quantities leftward (rightward) of
the contact discontinuity, and subscript $f$ denotes quantities inside 
the front.  For simplicity, we shall restrict our analysis to 
the case in which the colliding fluids have the same proper density,
pressure, and magnetic field, viz., $n_{+}=n_{-}$, 
$p_{+}=p_{-}$, $b{+}=b{-}$.  In this case a symmetric front will 
be created, i.e., $n_{f+}=n_{f-}$, $T^{\mu\nu}_{f+}=T^{\mu\nu}_{f-}$.
The problem is then characterized by seven independent variables: $T_f^{oo}$,
$T_f^{ox}$, $N_f$, $B_f$, $\beta_{s+}$, $\beta_{s-}$, $\beta_{c}$, where
$\beta_{s\pm}$ are the corresponding shock velocities, and
$\beta_c$ is the front velocity (i.e., the velocity of the contact 
discontinuity).  Since the proper energy density of the fast and slow
fluids is the same, the energy flux of the fast outflow, as measured in 
the injection frame, is larger than that of the slow outflow by roughly 
a factor of $(\Gamma_{-}/\Gamma_{+})^2$.  Thus, this case corresponds to a 
situation wherein the central engine undergoes an outburst during 
which energy is being injected into the system.  An 
alternative possibility, which will not be considered here, is that the 
fluid parameters change in such a way as to sustain the associated 
energy flux unchanged.  This would lead to the creation of an asymmetric 
front, which in some circumstances can give rise to a somewhat 
higher radiative efficiency.  To simplify the analysis further, we 
suppose that the proper density, pressure and magnetic field in the 
front are homogeneous, that $d_{+}=d_{-}=d$, and that the magnetic 
field dissipation rate
is proportional to the injection frame magnetic field, $B_f$,
and inversely proportional to the front length $\Delta X$, 
viz., $Q_b\Delta X=-\alpha_b B_f$, where $\alpha_b$ is a 
constant.  The latter assumption is not crucial, but is convenient since
it admits steady-state solutions in the adiabatic case (i.e., in the 
absence of radiative losses) which simplifies the choice of initial 
conditions (cf \S3.5).  As we have verified, 
the essential features of the solution are insensitive to the choice of 
the form of the term associated with magnetic field dissipation.  With 
the above simplifications eqs. (\ref{eq: app-front}) reduce to
		
\begin{eqnarray}
\label{eq: cont.}
\Delta X\frac{\partial}{\partial x^o} (N_{f})=\Delta X Q
- N_f(\beta_{s+}-\beta_{s-})+ N_{+}(\beta_{s+}-\beta_{+})
-N_{-}(\beta_{s-}-\beta_{-}),\\
\label{eq: B}
\Delta X\frac{\partial}{\partial x^o} (B_{f})=-\alpha_b B_{f}
- B_f(\beta_{s+}-\beta_{s-})+ B_{+}(\beta_{s+}-\beta_{+})
-B_{-}(\beta_{s-}-\beta_{-}),\\
\label{eq: Too}
\Delta X\frac{\partial}{\partial x^o} T^{00}_{f} =\Delta X S^0
- T_f^{00}(\beta_{s+}-\beta_{s-})+ (T_{+}^{00}\beta_{s+}-T_{+}^{0x})
-(T_{-}^{00}\beta_{s-}-T_{-}^{0x}),\\
\label{eq: Tox}
\Delta X\frac{\partial}{\partial x^o} T^{0x}_{f} =\Delta X S^x
- T_f^{0x}(\beta_{s+}-\beta_{s-})+ (T_{+}^{0x}\beta_{s+}-T_{+}^{xx})
-(T_{-}^{0x}\beta_{s-}-T_{-}^{xx}),
\end{eqnarray}
where $x^o=ct$, $t$ being the time measured in the injection 
frame, and where the shock and front velocities obey the 
algebraic equations,
\begin{eqnarray}
\label{eq:bs1}
(T^{oo}_f-T^{oo}_{+})\beta_{s+}+(T^{oo}_f-T^{oo}_{-})\beta_{s-}=
2T^{ox}_f-T^{ox}_{+}-T^{ox}_{-},\\
\label{eq:bs2}
(T^{ox}_f-T^{ox}_{+})\beta_{s+}+(T^{ox}_f-T^{ox}_{-})\beta_{s-}=
2T^{xx}_f-T^{xx}_{+}-T^{xx}_{-},\\
\label{eq:bc}
(\Gamma-1)T^{ox}_f\beta_c^2-[\Gamma(T^{oo}_f+b_f^2/2)-(\Gamma-1)
(r_f+b_f^2)]\beta_c +T^{ox}_f=0.
\end{eqnarray}
The last two terms on the RHS of equations (\ref{eq: cont.})- 
(\ref{eq: Tox}) account for the flux of the corresponding 
quantity incident into the front through the shock surfaces,
whereas the second term on the RHS describes the rate of change 
of the corresponding front quantity due to the front expansion. 
Finally, the expansion of the front is governed by the equation,
\begin{equation}
\frac{\partial}{\partial x^o}\Delta X=\beta_{s+}-\beta_{s-}, 
\label{eq: DX}
\end{equation}
and the crossing times of the forward and reverse shocks, denoted 
by $t_{R\pm}$, are given implicitly by
\begin{equation}
d_{\pm}/c=\pm\int_{0}^{t_{R\pm}}{(\beta_{s\pm}-\beta_{\pm})dt}.
\label{eq: tR}
\end{equation}
As discussed earlier, the above equations are valid only for times shorter 
than min \{$t_{R+}$,$t_{R-}$\}.  The equations describing the evolution of
the system at later times can be derived in a similar manner.
Formally, when $t>t_{R+}$ we replace $\beta_{s+}$ 
by $\beta_c$ in eqs. (\ref{eq: cont.})-(\ref{eq:bc})
, and set $N_{+}$, $B_{+}$, and $T^{\mu \nu}_{+}$, to zero.
Likewise, when $t>t_{R-}$ we set $\beta_{s-}=\beta_c$, 
$N_{-}=B_{-}=T^{\mu \nu}_{-}=0$.  In doing so we ignore the rare-faction 
waves that will be produced at the edges of the slabs due to the pressure
gradient there.
 											
\subsection{Kinetic equations}

Let $f_e(p_e,x)$ and $f_{\gamma}(p_{\gamma},x)$ denote the distribution 
functions of electrons (we do not distinguish electrons from positrons,
designating both by subscript $e$) and gamma-rays as measured in the 
injection frame, respectively.  The 
corresponding Boltzmann equations can be written as
\begin{equation}
p_e^{\mu}\partial_{\mu} f_e= C_e(f,p,x)+C_{inj},
\label{eq: Bolz-e}
\end{equation}
and

\begin{equation}
p_{\gamma}^{\mu}\partial_{\mu} f_{\gamma}= C_{\gamma}(f,p,x).
\label{eq: Bolz-g}
\end{equation}
The operators $C_e$ and $C_{\gamma}$ on the RHS of the above equations 
represent the interaction between pairs and gamma-rays, and are given 
explicitly in BL95.  They are restricted to the 
condition
\begin{equation}
\label{eq: E(Ce+Cg)}
\int{(C_{e}+C_{\gamma})d^3p}=0,  
\end{equation}
by virtue of energy conservation.  
The injection operator, $C_{inj}$, represents the change in energy and 
momentum of electrons (positrons) due to their interaction with 
the large scale electromagnetic field; that is, as a result of shock
acceleration, magnetic reconnection, or stochastic acceleration owing to
absorption of nonlinear plasma waves. 
Since the injection operator preserves the total electron number density,
it must satisfy,
\begin{equation}
\int{C_{inj}\frac{d^3p}{p^o}}=0.
\label{eq: Cinj}
\end{equation}

The angular distribution of pairs and gamma-rays is expected to be strongly 
beamed along the direction of propagation of the front in the injection 
frame.  Thus, we can approximate the distribution functions as: $f_{\alpha}
=p_{\alpha}^{-2}n_{\alpha}(E_{\alpha},x)\delta(\mu_{\alpha})$; $\alpha
= (e,\gamma)$, with $\mu_{\alpha}$ being the cosine of the angle between the
momentum of a species $\alpha$ and the flow velocity, and 
$n_{\alpha}(E_{\alpha})$ the corresponding number density of a species 
$\alpha$ per unit energy.  Integrating eqs. (\ref{eq: Bolz-e}) and
(\ref{eq: Bolz-g}) over a region encompassing the front in the manner 
described in the appendix, assuming again that the distribution functions are 
homogeneous inside the front, and averaging over $\mu_{\alpha}$ 
yields to a good approximation,
											
\begin{eqnarray}
\label{eq: ne}
\Delta X\frac{\partial}{\partial x^o}n_{ef}=
\int{(C_e+C_{inj})dx}
-n_{ef}(\beta_{s+}-\beta_{s-})
+n_{e+}(\beta_{s+}-1)-n_{e-}(\beta_{s-}-1),\\ 
\label{eq: ng}
\Delta X\frac{\partial}{\partial x^o}n_{\gamma f}=
\int{ C_{\gamma}dx}
-n_{\gamma f}(\beta_{s+}-\beta_{s-})
+n_{\gamma +}(\beta_{s+}-1)-n_{\gamma -}
(\beta_{s-}-1).
\end{eqnarray}
											
Now, the electron density in the fluids exterior to the front, viz., 
$n_{e+}$ and $n{e-}$, are given as input, and can be absorbed 
into the definition of
the injection operator.  For convenience we  redefine the injection
operator as follows: $\int{C_{inj} dx}\rightarrow \int{\bar{C}_{inj}dx}
=\int{C_{inj}dx}+n_{e+}(\beta_{s+}-1)-n_{e-}(\beta_{s-}-1)$.  By employing
 eq. (\ref{eq: Cinj}) we arrive at,
\begin{equation}
\int{dx}\int{\bar{C}_{inj}dE_e}=N_{+}(\beta_{s+}-\beta_{+})
-N_{-}(\beta_{s-}-\beta_{-}).
\label{eq: Cinj-bar}
\end{equation}
This condition simply reflects the fact that particle acceleration 
preserves the total number density of particles incident into the front.
A second constraint on $\bar{C}_{inj}$ is obtained by multiplying
eq. (\ref{eq: ne}) by $E_e$, and then integrating over $dE_e$:
\begin{equation}
\int{dx}\int E_e{\bar{C}_{inj}dE_e}=T^{oo}_{sf}(\beta_{s+}-\beta_{s-})
+\int{\frac{\partial T^{oo}_{sf}}{\partial x^o} dx}-\int{S^o dx}.
\label{eq: dissp-eng}
\end{equation}
where $T^{oo}_{sf}$ is given by eq. (\ref{eq: Tu}).  Note that 
the RHS of the last equation equals
the fraction of total energy flux dissipated inside the front.

We must also determine $n_{\gamma+}$ and $n_{\gamma-}$.  Under the
approximation of perfect beaming invoked above $n_{\gamma-}=0$ and
$n_{\gamma+}=n_{\gamma f}$, since the only source of beamed radiation is 
the front itself.  With the above results eq. (\ref{eq: ng})
simplifies to,
\begin{equation}
\label{eq: ng2}
\frac{\partial}{\partial x^o}n_{\gamma f}=
C_{\gamma}+(\Delta X)^{-1}n_{\gamma f}(\beta_{s-}-1).
\end{equation}
Again, we assume that for $t>t_{R\pm}$ the time change of the distribution 
functions is dictated by eqs. (\ref{eq: ne}) and (\ref{eq: ng2}) with 
$\beta_{s-}$ and $\beta_{s+}$ replaced by $\beta_c$ and $n_{e\pm}$
set to zero.   

The emergent gamma-ray spectrum depends on the pair production opacity
contributed by the ambient radiation field ahead of the front.  
Gamma-rays for which the pair production optical depth to infinity
largely exceeds unity, will be converted into lower energy 
pairs and gamma-ray via pair cascades upstream the forward shock. 
(We note that the escape probability of gamma-rays at a given energy 
from the front may be considerably different than the probability that 
they will escape the system to infinity.  In fact, the pair production 
optical depth of the front itself will be smaller than unity as 
long as its axial length is smaller than the corresponding 
gamma-spheric radius.)  This leads, as shown below, to a strong 
suppression of the emitted flux at the corresponding  energies.  
The calculation of the emitted spectrum proceeds
as follows:  At every time step we integrate the equations describing the
spatial evolution of the cascade,
\begin{eqnarray}
\label{eq: nesce}
\frac{\partial}{\partial \ln r}n_{\gamma}(E_{\gamma})=C_{\gamma},\\
\frac{\partial}{\partial \ln r}n_{e}(E_e)=C_{e},
\label{eq: nescg}
\end{eqnarray}
starting at the current position of the forward shock, 
$r(t)=r_o+\int{\beta_{s+}dt}$, and subject to the boundary 
conditions $n_e[r(t),E_e]=n_{e+}$, $n_{\gamma}[r(t),E_{\gamma}]=
n_{\gamma f}(t,E_{\gamma})$.  Here $n_{\gamma}(r,E_{\gamma})$ 
is the number density per unit energy of escaping gamma-rays ahead 
of the forward shock.  The assumption underline these calculations 
is that the cascade develops on a sufficiently short time scale to 
render any retardation effects negligible.  We also ignore
possible alterations of the front structure due to momentum exchange 
between emitted gamma-rays and the slow outflow (i.e., upstream of the 
forward shock).  The inclusion of this effect complicates the numerics
substantially and is beyond the 
scope of our analysis.  We anticipate, though, that such alterations of 
the front structure will not affect the emission characteristics 
significantly.

\subsection{Determination of the source functions}

The source term describing the rate of change of the electron density 
in the front due to formation of pair cascades, is
determined by integrating eq. (\ref{eq: ne}) over energy, and using eqs. 
(\ref{eq: Cinj-bar}) and (\ref{eq: Nu}). 
One then obtains, 

\begin{equation}
Q=2\int{C_e dE_e}=2\int{\kappa_{pp}(E_{\gamma})n_{\gamma f} dE_{\gamma}},
\label{eq: Q}
\end{equation}
where $\kappa_{pp}(x,E_{\gamma})$ is the pair production opacity, and the
second equality has been obtained using eq. (4.12) of BL95.
Likewise, the radiative energy loss rate is obtained by taking the 
first moment of
eq. (\ref{eq: ne}), i.e., multiplying the equation by $E_e$ and then 
integrating over $dE_e$, and by using eqs. (\ref{eq: dissp-eng}) and 
(\ref{eq: Tu}):
\begin{equation}
S^o=\int{E_e C_e dE_e}.
\label{eq: S}
\end{equation}
Under the assumption of perfect beaming the momentum loss rate equals
the energy loss rate, viz., $S^x=cS^o$.  By employing eqs. (\ref{eq: 
E(Ce+Cg)}) and (\ref{eq: ng2}) we can rewrite
the rates associated with radiative losses in the form
\begin{equation}
S^x/c=S^o=-\frac{\partial T^{oo}_{\gamma}}{\partial x^o}
-(\Delta X)^{-1}T^{oo}_{\gamma}(1-\beta_{s-}),
\label{eq: Sx}
\end{equation}  
where $T^{oo}_{\gamma}=\int{E_{\gamma}n_{\gamma}d\ln E_{\gamma}}$ is the 
gamma-ray energy density.  Note that the latter equation is essentially 
the energy equation for the system of gamma-rays.   
\subsection{The injection operator}						

As stated above, the injection operator represents the energy 
redistribution of electrons (positrons) resulting from their interaction 
with the electromagnetic field, e.g., due to Fermi acceleration, magnetic
reconnection, or absorption of nonlinear plasma waves excited by the 
shock.  The physics of these processes is yet to be understood better
before $C_{inj}$ can be derived from first principles.  Here we settle for
a simple prescription in which a small fraction of 
the pairs incident into the front are injected to some maximum energy,
assumed to be fixed in the observer frame, and 
the rest are redistributed at much lower energies.
The maximum injection energy, $E_{emax}$, and the fraction $\eta$ 
of the dissipation energy that is being 
injected to $E_{emax}$ are treated as free parameters.  To be concrete,
we choose 
\begin{equation}
\bar{C}_{inj}=n_1\delta(E_e-E_o)+n_2\delta(E_e-E_{emax})
\label{eq:cinj}
\end{equation}
with $n_1$, $n_2$ and $E_o$ determined from eqs. (\ref{eq: Cinj-bar}) and 
(\ref{eq: dissp-eng}), and the requirement 
that $n_2E_{emax}$ comprises a fraction $\eta$ of the total dissipation 
energy.  Other forms of $\bar{C}_{inj}$ can be readily 
derived, for example, a thermal
distribution with a power law tail.  We find though that the results are
insensitive to the exact form of the injection operator provided that 
particle acceleration is sufficiently efficient, in the sense that a 
significant fraction of the injection energy is carried by electrons having
energies near $E_{emax}$.  It should be noted that the assumption that 
$E_{emax}$ is fixed in the observer frame is unrealistic.  A more realistic 
prescription would be to take $E_{emax}$ to be fixed in the rest frame
of the front.  Unfortunately, this complicates the numerics considerably.
However, variations of $E_{emax}$ due to the changing front velocity 
should not affect significantly the evolution of the gamma-ray 
distribution function at sufficiently low energies. 
											
\subsection{Initial condition}

In order to integrate the above equations, one must specify the initial 
position and structure of the front, and the initial energy distribution of 
electrons.  A simple and convenient choice would be to use the structure 
of an adiabatic front.  Below, we employ the front structure calculated by 
Levinson \& Van Putten (1997).  Specifically, we first solve 
algebraically eqs. (\ref{eq: cont.})-(\ref{eq:bc}) 
in the absence of radiative losses (i.e., with $Q=S^o=S^x=\alpha_b=0$),
and assuming steady state ($\partial/\partial x^o=0$),
for the chosen input parameters.  The front quantities 
thereby obtained then serve as initial values for the integration of
the radiative front equations.  This choice is viable
if the creation of the front occurs on a timescale much shorter than the 
radiative cooling time.  An initial electron distribution subject to
the restrictions that the number density and total energy (temperature) 
equal those computed for the adiabatic front is also specified.
The initial thickness of the front, $\Delta X(t=0)$, is taken 
to be small enough, so that the energy loss 
rate due to cooling of the initial population 
of electrons is well below the rate of energy deposition in the front.
This renders the results highly insensitive to the choice of initial 
electron distribution. 

\section{Results} 

Equations (\ref{eq: cont.})-(\ref{eq: tR}), (\ref{eq: ne}), (\ref{eq: ng2})-
(\ref{eq: nescg}) and (\ref{eq:cinj}) have been integrated 
numerically using the initial conditions discussed in \S 3.5. 
The source functions $Q$ and $S^o$ have been computed using eqs. 
(\ref{eq: Q}) and (\ref{eq: S}). 
We have verified that the solution is indeed highly 
insensitive to the choice initial electron distribution when 
the initial length of the front is sufficiently small, as stated
in \S 3.5.  The equations have been modified in a manner described in 
\S 3.1 after each shock crossing 
(first at $t=$ min $\{t_{R+},t_{R-}\}$ and again at 
$t=$ max $\{t_{R+},t_{R-}\}$), and the integration 
continued using the modified equations.  In the following examples, 
the Lorentz factors of the slow and fast fluids and the rest frame 
Alf\'ven 4-velocity have been taken to be 
5, 20, and 10, respectively, and the initial electron distribution 
has been taken to be Maxwellian with appropriate temperature and density.
A rapid magnetic field dissipation has been 
invoked with $\alpha_b=0.5$.
The standard soft photon intensity (consisting of a broken power 
law with a steeper slope below about 0.5 KeV) defined in 
Levinson \& Blandford (1995) has been adopted for the calculations.
To be concrete, we used the following form for the intensity of 
external radiation:
$$
I_s(E_s,t)=\frac{\epsilon L_s}{4\pi r^2(t)}g(E_s);\ \ \ \ \ 10^{-5}<E_s
<0.1,
$$
with $r(t)=r_o+\int{\beta_{c}dx^o}$, and $g(E_s)\propto (E_s/E_o)^{-1/2}[
1+(E_s/E_o)^{-1}]$; $\int g(E_s) dE_s=1$.  Here $E_o$ is the break energy and
is taken to be 0.5 keV, and $\epsilon<<1$ is the fraction of nuclear luminosity
that is reprocessed or scattered by surrounding gas. 	
As a check, we integrated eq. (\ref{eq: Sx}) to obtain $T^{oo}_{\gamma}$ and 
compared the result with total flux emitted (i.e., $\int{E_{\gamma}
n_{\gamma}d\ln E_{\gamma}}$) at every time step.  The agreement was 
typically better than 2\%.
									
To illustrate the dynamics of the system we first consider the 
limit of continuous ejection; that is, $d/r_o\rightarrow\infty$.
The time evolution of the front quantities is depicted in fig. 1, 
where the 4-velocity of the contact discontinuity and the 
two shocks (upper left panel), the proper density (upper right panel), 
the total pressure (bottom left panel), and the proper magnetic 
pressure (bottom right panel)
are plotted against log$(ct/r_o)$, $t$ being the injection frame 
time, for $d/r_o=100$, $(\epsilon L_s)_{45}/r_{16}=1$ (solid lines) and 
$(\epsilon L_s)_{45}/r_{16}=10^{-2}$ (dashed lines).    
Here $(\epsilon L_s)_{45}$ is the scattered luminosity in units
of 10$^{45}$ erg s$^{-1}$, and $r_{16}=r_o/(10^{16} {\rm cm})$. 
The corresponding radiative efficiency, defined as the fraction of 
dissipation power radiated by the front, and given explicitly by,
$$    
\Phi(t)=\frac{\int{n_{\gamma }(E_{\gamma},t)
(1-\beta_{s+})dE_{\gamma}}}{\int{dx}\int E_e{\bar{C}_{inj}dE_e}},
$$
is exhibited in fig. 2.  As seen from figs. 1 and 2, the 
front velocity and expansion rate, 
initially equal those of an adiabatic front, decrease as the emitted 
energy flux rises.  The rest mass density and total pressure increase
correspondingly.  After the peak emission is
reached, the front starts accelerating and adiabatic cooling 
becomes gradually more important until, ultimately, the initial 
structure and velocity of the front are restored.  The increase in number
density is partly due to pair production and partly due to enhanced 
compression resulting from the drop in expansion rate.  The change in 
total pressure is primarily due to magnetic field compression.
Quite generally, we find that when $t_{R\pm}>l/c$, the maximum of the 
flux is reached at a distance $l$ from the creation radius $r_o$, where 
$l$ is the gradient length scale of the intensity of ambient 
radiation at $r_o$ ($l\sim r_o$ in this example), and the duration of 
the flare corresponds to several times $l$ (see below).  When 
$t_{R\pm}<l/c$ the timescale of the flare is determined by the 
shock crossing times.  The evolution of the system in this case 
is essentially the same as described above for times shorter than 
$t_{R\pm}$.  At later times, however, energy supply to the hot 
outflow terminates, and the outflow begins to 
cool radiatively and decelerate until radiative losses become small.  
The radiative efficiency, namely the ratio of radiated and 
dissipated energy fluxes increases with increasing values of 
$(\epsilon L_s)_{45}/r_{16}$, as seen from fig. 2.  It approaches 
80 percent near the peak in this example 
for $(\epsilon L_s)_{45}/r_{16}=1$, and is generally around this value  
for typical parameters of gamma-ray blazars.    
		
The rate of energy dissipation in the front is presented in fig. 3, and
it is seen that it increases as the radiated flux increases.  The reason
is that radiative losses lead to deceleration of the front and shocks 
(see fig. 1) and, consequently, enhancement in the rate of energy deposition
behind the shocks, particularly the reverse shock, as can be 
inferred from eq. (\ref{eq: Too}).  This positive feedback has 
important implications
for the emitted spectrum, particularly in the presence of large pair
production opacity, that are discussed below.   
Fig. 4 depicts typical light curves computed for $d/r_o>1$ and 
different values of $(\epsilon L_s)_{45}/r_{16}$.
In this figure the total apparent gamma-ray luminosity is plotted 
against the time measured by a distant observer, 
$t_{obs}=\int_{0}^{t}{[1-\beta_{s+}(t')
\cos\theta]dt'}$, where $\beta_{s+}$ is the velocity of 
the forward shock, and $\theta$ is the angle to the line of sight ($\theta=0$
in this example).  The strong dependence of the flare intensity on
the parameter $(\epsilon L_s)_{45}/r_{16}$ is evident.  This is 
one consequence of the positive feedback mentioned above. 
The decay of the flare in this regime depends on the 
radial variation of the intensity of background radiation 
and the expansion rate of the front, and is typically longer 
and more gradual than the rise.  Further, the 
rise time increases slightly with increasing $L_{45}/r_{16}$.  The 
reason is that larger opacity results in smaller shock and 
front velocities during peak emission (see fig. 1) and, 
therefore, smaller beaming factor.  From fig. 4. it is seen 
that the rise time is of order $(l/2c)\Gamma_{s+}^{-2}(t_{peak})
\simeq10^{-2}r_o/c$, where $\Gamma_{s+}(t_{peak})$ is the Lorentz
factor of the forward shock near maximum flux,
and that the flare 
duration is several times longer.  For $r_o$ between 10$^{15}$
and 10$^{18}$ cm (cf. \S 2) this corresponds to a rise time in 
the range between several minutes and several weeks.  However, 
gamma-rays of a given energy 
cannot escape from a radius smaller than their gamma-spheric radius, 
so that the duration of a flare in a given band is also
limited by the pair production opacity.  In general, we find that  
for parameters typical to the powerful blazars 
the flare duration in the EGRET band can range from several hours to 
several weeks for small viewing angles (i.e., $\theta<\Gamma^{-1}$), when
$d/r_o>1$.  The shape of the flare is substantially altered 
when $d/r_o$ becomes sufficiently small.  This is demonstrated in 
Fig. 5 where light curves computed for $L_{45}/r_{16}=1$ and different values
of $d/r_o$ are displayed.  As seen, for values of $d/r_o$ 
smaller than unity the time scale of the flare, as measured by a distant
observer, is of order $d/c$; the light curves exhibit a 
steep decline at times $t_{obs}>d/c$.  For sufficiently thin 
slabs ($d/r_o<2\times10^{-2}$ in this example) the decay time of 
the flare is comparable to the rise time or even slightly shorter.     
 
In order to examine the relationship between the emergent fluxes
in different high-energy bands predicted by our 
model, we divided the energy interval into several subintervals
and followed the time evolution of the flux in each band.
Quite generally we find that i) only a relatively 
small fraction of the radiated energy is emitted above 
the gamma-spheric energy at which the pair-production optical depth
at $r\sim r_o+l$ (the front position at peak emission) equals unity,
ii) the time of peak emission and flare duration increase
with increasing gamma-ray energy in this energy range, 
and iii) below that energy variations in the fluxes at 
different X-and gamma-ray bands occurs roughly simultaneously, but not 
necessarily with the same amplitude.  This behavior is 
also a consequence of the positive feedback discussed above.
An example is given in fig. 6, where the time evolution 
of the energy fluxes emitted in four equally spaced subintervals in 
the energy interval 5 MeV to 50 GeV, for $d/l>1$ is 
presented.  Similar behavior has been found in the case $d/l<<1$. 
This result is in contrast with the simultaneous flaring predicted
by the model of RL97, which does not take into 
account gamma-ray attenuation by pair production on external photons.   

We also checked the dependence of the front dynamics and emission on the 
rate of magnetic field dissipation.  We find that
as $U_A$ increases above unity, the front expands more rapidly, adiabatic
cooling becomes more important and, hence, the radiative efficiency drops.
The rapid expansion results in a slower decay of the flare, but 
otherwise the shape of the light curve is not altered significantly.

\section{Conclusions} 

This paper considers the production of gamma-ray flares through inverse
Compton scattering of external radiation by pairs accelerated behind 
internal shocks, that are driven by temporal fluctuations in the 
parameters of a magnetized, relativistic outflow.  A self-consistent 
model capable of describing the dynamics of the front and the time 
evolution of the angle averaged pair and gamma-ray distribution 
functions, has been developed and employed to calculate gamma-ray 
light curves in blazars.

The main results and conclusions of this study are:

1)  The shape and timescale of the flare depend on the 
ratio of the thickness of ejected fluid slab, $d$, and
the gradient length scale of the background radiation intensity 
at the radius of shock formation, $l$.  When $d/l$ is 
sufficiently small, such that the propagation time of the shocks across
the fluid slab is shorter than $l/c$, the flare duration, as measured 
by an observer at small viewing angle, is $\sim d/c$, and the 
shape of the light curve is 
roughly symmetric.  For larger values of $d/l$ the flare reaches
its maximum at a distance $l$ from the creation radius.  It then 
decays gradually until time $d/c$, after which the flux 
declines steeply.  For $d/l>1$ the shape of the flare is determined 
essentially by the radial variations of the intensity of ambient 
radiation and the expansion rate of the front; typically, the rise 
is fast compared with the decay.  The above results suggest that  
different types of flares will
be produced in different sources, or even in the same source 
under different conditions.  This expectation appears to be 
consistent with the variety of types of EGRET flares (e.g., 
Hartman et al. 1993; Kniffen et al. 1993; Mattox et al. 1997) 
observed in gamma-ray blazars.  In the event of impulsive outflow 
ejection (i.e., ejection time of order the dynamical time, $\tau_{inj}\simeq 
\tau_{acc}$; cf. \S 2) the axial length of the fluid slab can be
as small as the size of the central engine.  If the latter is associated
with the gravitational radius of the putative black hole, $r_g$, then
flare durations as short as $r_g/c\sim 50[M_{BH}/(10^7M_{\odot})]$ 
sec are plausible.  Successive ejection of jets with short duty
cycle may also lead to a substructure in the light curve of a longer
duration outburst (as often seen in GRBs).  In general we 
find that for parameters typical to extragalactic jets, namely $\Gamma$ 
of order 10 and comoving Alf\'ven 4-velocity between 0 and 10, the flare 
timescale can lie in the range between several minutes and several weeks
(for small viewing angles).  
We note that the shape of the
light curve may be significantly altered when the SSC process 
becomes important (cf. RL97).  The study of SSC flares is left for future 
investigation.

2) The peak flux and flare intensity depend sensitively on the 
Thomson opacity near $r_o$, specifically, on the value of $L_s/r_o$,
where $L_s$ is the fraction of nuclear luminosity that is reprocessed
or scattered across the jet. 
This behavior is a consequence of a positive feedback that gives rise 
to a strong enhancement of the energy deposition rate during peak emission. 
For parameters typical to the powerful gamma-ray blazars,
the radiative efficiency, namely the fraction of dissipation energy 
that is radiated by the front is typically high ($>60\%$).  

3)  The amplitude of variations of the flux emitted at energies  
for which the pair-production optical depth is initially larger
than unity is much smaller than that at lower energies.  The 
flare, in this energy interval, propagates from low to high energies
and its duration increases with energy.
At lower energies, flaring in different X-and gamma-ray bands
occurs roughly simultaneously, but with possibly different amplitudes,
depending on the spectra of ambient radiation and injected electrons.
The claimed invariance of the gamma-ray spectrum during the strong EGRET flare 
observed in PKS1622-297 (Mattox et al. 1997) is not in conflict with that
model prediction, since the sensitivity was insufficient to resolve any 
time lags, if present, between the fluxes in the two
energy subintervals (below and above 300 MeV) analyzed by those authors,
particularly in view of the small amplitudes anticipated at
high EGRET energies.  Despite the small amplitudes, it may be possible 
to observe time lags between the emission at hard (but still well below
TeV where absorption by the IR background is strong) and soft 
gamma-ray energies with a next generation gamma-ray telescope.  
The detection of such lags is an important test for this model.
One implication of the above results is that in some cases strong flares
may be observed in some energy bands while the simultaneous variations 
in other bands may appear modest or even small.
The simultaneous X-ray/TeV flare
reported recently for Mrk 421 and the lack of significant variations
of the EGRET flux (Macomb, et al. 1995; Takahashi et al. 1996) is 
an example.  The hard 
X-ray flux may be due to either inverse Compton emission of the thermal 
electrons or synchrotron emission of the highest energy electrons.  
Alternatively, this event may be explained as an SSC flare 
resulting from changes in the tail of the distribution of 
non-thermal electrons (Takahashi et al. 1996).  

I thank the referee, R.V.E. Lovelace, for useful comments.
Support by Alon fellowship and a TAU Research Authority grant 
is acknowledged.
 
\appendix
\section{Appendix: Derivation of the front equations}

Equations (\ref{eq: MHD}) are of the form,
\begin{equation}
\partial_{\mu}A^{\mu}=B.
\label{eq: a1}
\end{equation}
Integrating the above equation over a region in
space-time enclosed by the world lines associated with the left
shock, the contact discontinuity and the lines defined by the 
equations $t=t_o$; $t=t_o+\Delta t$, where $t_o$ is some fiducial 
time [see fig. (7)] , and using the generalized Gauss's theorem,
\begin{equation}
\int{A^{\mu}d^3\Sigma_{\mu}}=\int{B d^4\Omega},
\end{equation} 
where $d^4\Omega$ is a volume element and $d^3\Sigma_{\mu}$ is the 
surface area of the 3d surface enclosing the region of integration yields,
\begin{eqnarray}
\int_{t_o}^{t_o+\Delta t}{\{A^{o}_{-}\beta_{s-}-A^x_{-}+A^{x}_{f-}(x_c)
-A^{o}_{f-}(x_c)\beta_{c}\}cdt}
+\int_{x_c(t_o)}^{x_{-}(t_o)}{A^{o}_{f-}(t_o)dx}\nonumber\\
-\int_{x_c(t_o+\Delta t)}^{x_{-}(t_o+\Delta t)}{A^{o}_{f-}(t_o+\Delta t)dx}=
\int_{t_o}^{t_o+\Delta t}{cdt\int_{x_c(t)}^{x_{-}(t)}{B(t)dx}},
\end{eqnarray}    
where it has been assumed that $A^{\mu}$ and $B$ depend solely on the 
axial coordinate (i.e., along the direction of motion).
In the limit $\Delta t\rightarrow0$, and using the expansions $x_c(t_o+dt)
\simeq x_c(t_o)+\beta_c(t_o)dt$;  $x_{-}(t_o+dt)
\simeq x_{-}(t_o)+\beta_{s-}(t_o)dt$, we arrive at,

\begin{equation}
A^{o}_{-}\beta_{s-}-A^x_{-}-A^{o}_{f-}(x_-)\beta_{s-}+A^{x}_{f-}(x_c)+
\int_{x_{-}}^{x_c}{\frac{\partial A_{f-}^o}{c\partial t}dx}-
\int_{x_{-}}^{x_{c}}{B_{-}dx}=0.
\end{equation}

Likewise, integrating eq. (\ref{eq: a1}) over a portion of space 
time containing 
the region rightward to the contact discontinuity yields,

\begin{equation}
-A^{o}_{+}\beta_{s+}+A^x_{+}+A^{o}_{f+}(x_+)\beta_{s+}-A^{x}_{f+}(x_c)+
\int_{x_{c}}^{x_{+}}{\frac{\partial A_{f+}^o}{c\partial t}dx}-
\int_{x_{c}}^{x_{+}}{B_{+}dx}=0.
\end{equation}
Under the assumption that $A_{f\pm}$ are homogeneous inside the front, (i.e.,
independent of the axial coordinate x, but may still have different values
on each side of the contact discontinuity), we obtain the following 
differential equations for $A_{f\pm}(t)$    
\begin{equation}
\pm(\Delta X_{\pm})^{-1}(A^{o}_{\pm}\beta_{s\pm}-A^x_{\pm}-
A^{o}_{f\pm}\beta_{s\pm}+A^{x}_{f\pm})-
\frac{\partial A_{f\pm}^o}{\partial x^o}+B_{\pm}=0,
\end{equation}
where $\Delta X_{\pm}(t)=\pm\int_{x_c(t)}^{x_{\pm}(t)}{dx}$ is the distance between 
the forward (reverse) shock and the contact discontinuity.

Substituting the front quantities for $A^{\mu}$ one obtains,
\begin{eqnarray}
\pm[N_{f\pm}(\beta_{s\pm}-\beta_c)-N_{\pm}(\beta_{s\pm}-\beta_{\pm})]
+\Delta X_{\pm}\left(\frac{\partial N_{f\pm}}{\partial x^o}-Q_{\pm}\right)=0,
\nonumber\\
\label{eq: app-front}
\pm[B_{f\pm}(\beta_{s\pm}-\beta_c)-B_{\pm}(\beta_{s\pm}-\beta_{\pm})]
+\Delta X_{\pm}\left(\frac{\partial B_{f\pm}}{\partial x^o}-Q_{b\pm}\right)=0,\\
\pm[(T^{00}_{f\pm}\beta_{s\pm}-T^{0x}_{f\pm})-(T^{00}_{\pm}\beta_{s\pm}-T^{0x}_{\pm})]
+\Delta X_{\pm}\left(\frac{\partial T^{00}_{f\pm}}{\partial x^o}-S^0_{\pm}\right)=0,
\nonumber\\
\pm[(T^{0x}_{f\pm}\beta_{s\pm}-T^{xx}_{f\pm})-(T^{0x}_{\pm}\beta_{s\pm}-T^{xx}_{\pm})]
+\Delta X_{\pm}\left(\frac{\partial T^{0x}_{f\pm}}{\partial x^o}-S^x_{\pm}\right)=0.
\nonumber
\end{eqnarray}
Here $B_{f\pm}=b_{f\pm}\gamma_c$ is the magnetic field as measured in the 
injection frame, $\beta_{s\pm}$ are the corresponding shock velocities,
and $\beta_c$ is the velocity of the contact discontinuity. 
											   
\break

\break

\begin{figure}
\vspace{10cm}  
\includegraphics{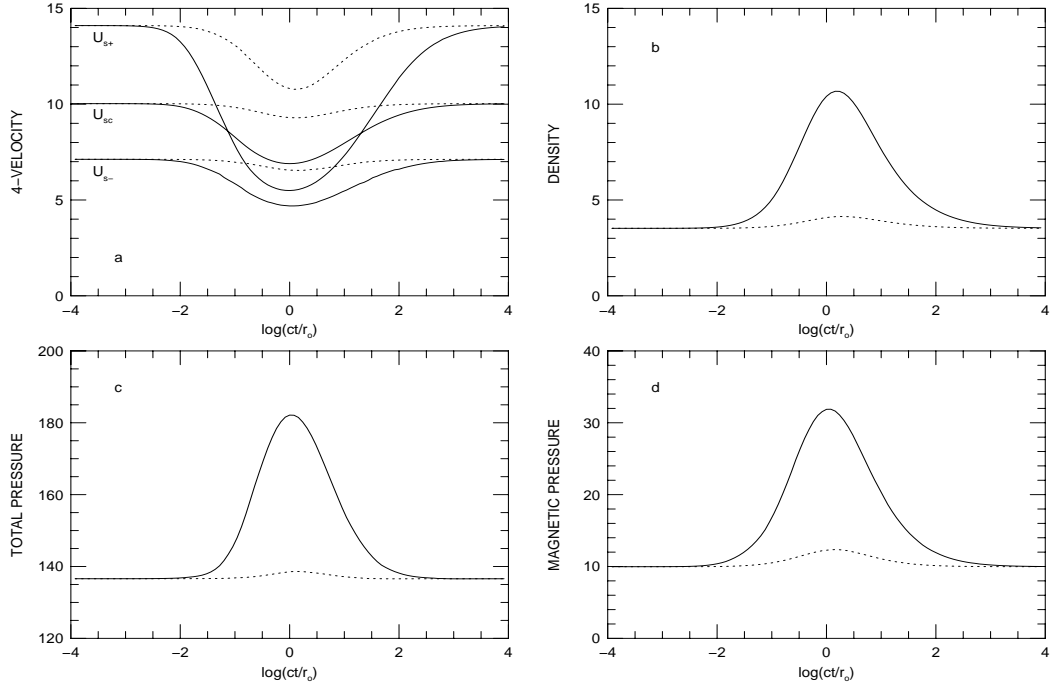}
\caption{Time evolution of front quantities.  Shown are (a) the 
4-velocity of the contact discontinuity (labeled $U_{c}$), forward shock 
(labeled $U_{s+}$) and reverse shock (labeled $U_{-}$), (b) the proper 
density, (c) the total pressure, and
(d) magnetic pressure as a function of ln($ct/r_o$), for  
$(\epsilon L_s)_{45}/r_{16}=1$ (solid line) and $10^{-2}$ (dotted line).
Here $(\epsilon L_s)_{45}$ is the fraction of ambient luminosity 
scattered across the jet in units of $10^{45}$ ergs s$^{-1}$,
and $r_{16}$ is the radius of shock formation in units of 10$^{16}$ cm.}
\end{figure}

\break	
\begin{figure}
\vspace{10cm}  
\includegraphics{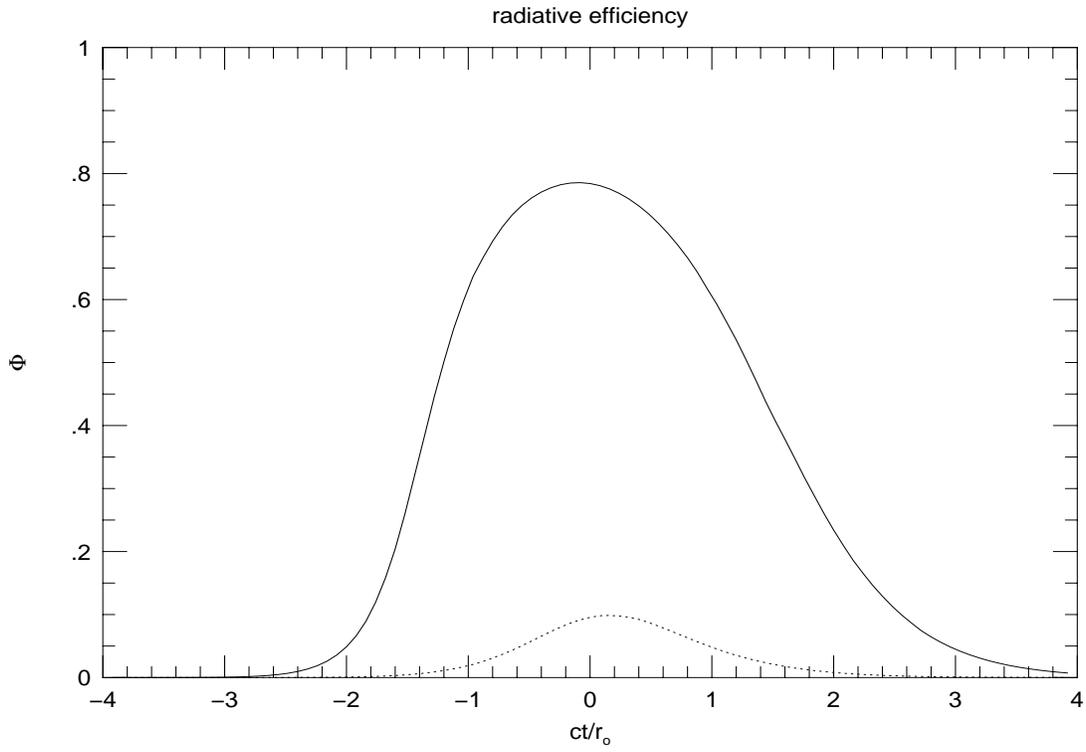}
\caption{Radiative efficiency, defined as the fraction 
of dissipation power radiated by the front (see text), versus 
log of injection frame time.}
\end{figure}

\begin{figure}
\vspace{10cm}  
\includegraphics{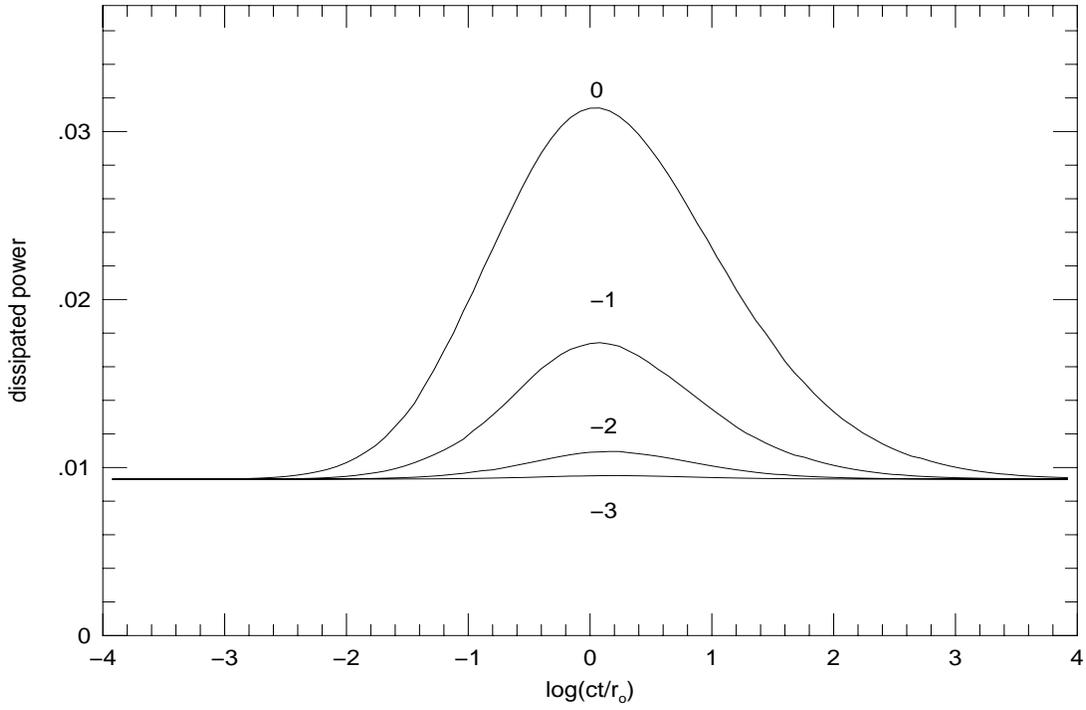}
\caption{Time evolution of the dissipation power.
The dissipation flux is normalized to the bulk energy flux of the
fast fluid.  Curves are labeled by values of 
log[$(\epsilon L_s)_{45}/r_{16}$]}
\end{figure}

\begin{figure}
\vspace{10cm}  
\includegraphics{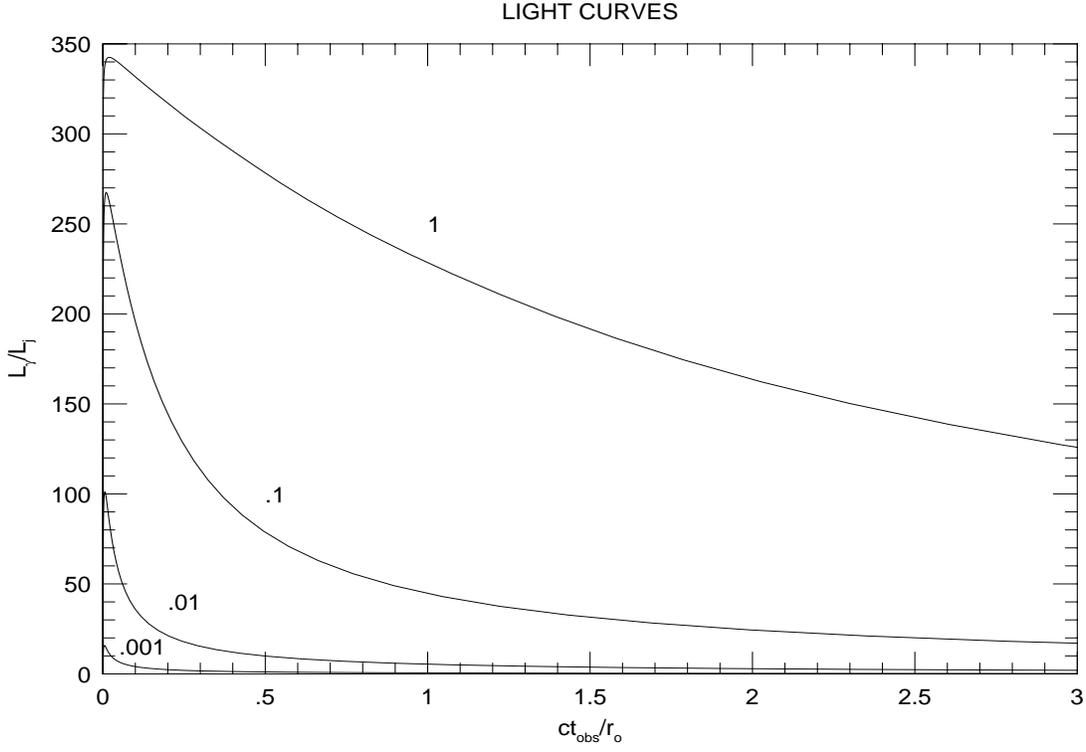}
\caption{Typical gamma-ray light curves produced by the 
model for $d/r_o>1$.  Shown is the total apparent luminosity 
in units of $L_j$, the power of the ejected (fast)
outflow, against $ct_{obs}/r_o$, $t_{obs}$ 
being the time measured by distant observer (see text for details).  The 
numbers that label the curves are the values of 
$(\epsilon L_s)_{45}/r_{16}$.}
\end{figure}

\begin{figure}
\vspace{10cm}  
\includegraphics{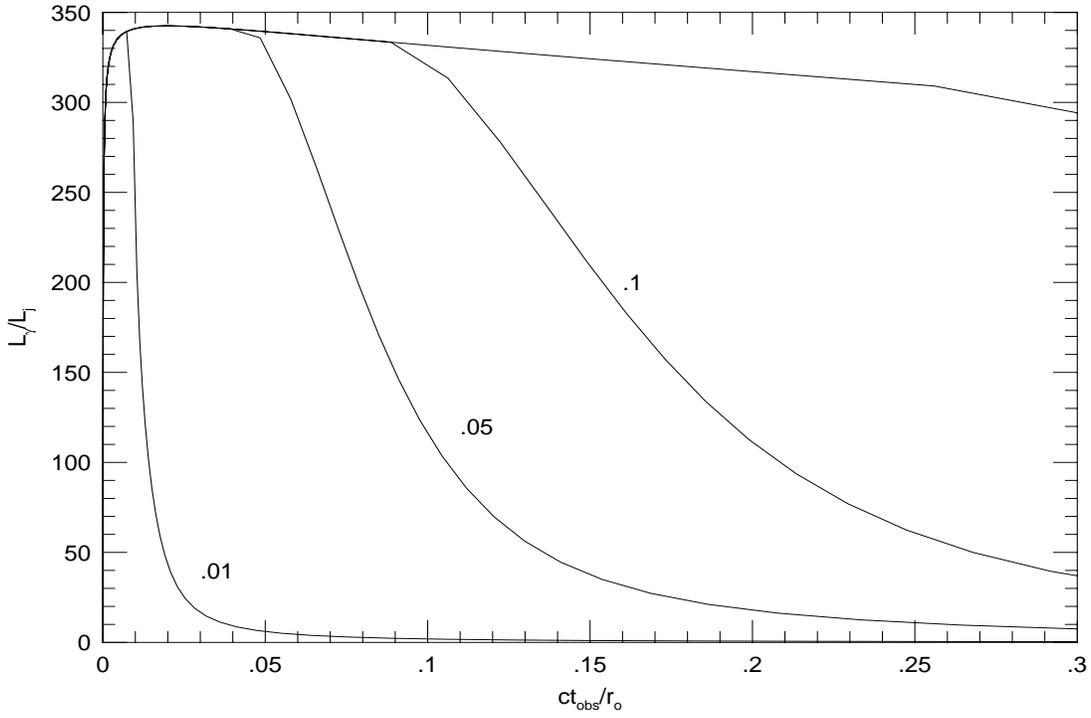}
\caption{The same as fig. 4, but for $(\epsilon L_s)_{45}/r_{16}=1$ and 
different values of $d/r_o$ (label the curves).}
\end{figure}

\begin{figure}
\vspace{10cm}  
\includegraphics{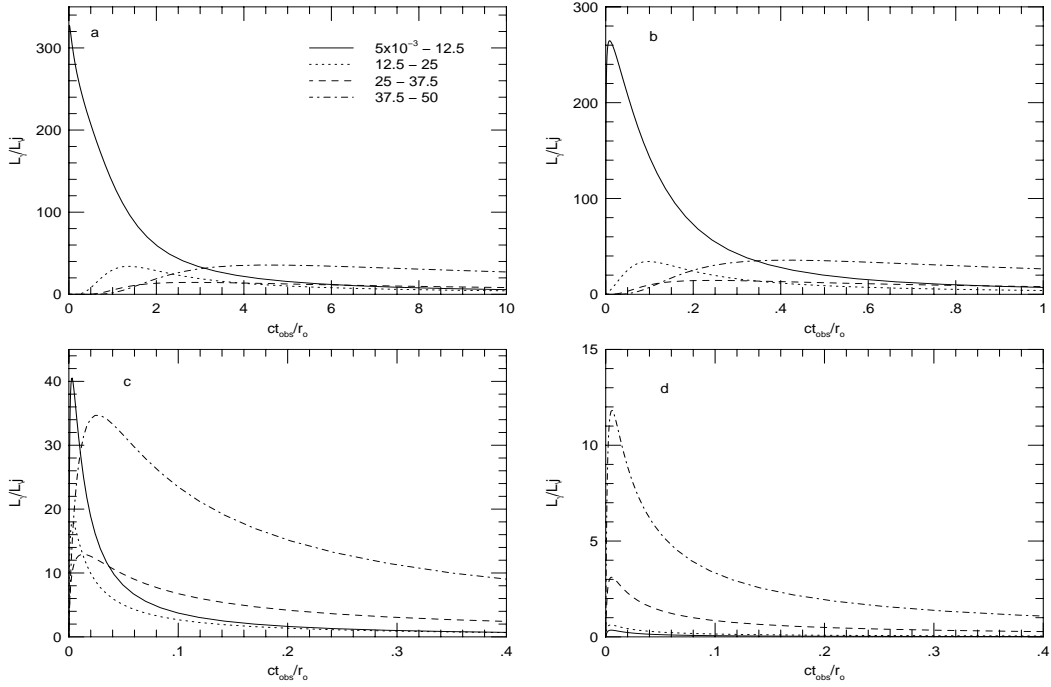}
\caption{Spectral evolution: 
displayed are the total apparent luminosities in four equally spaced energy 
intervals in the range 5 MeV to 50 GeV (indicated in window a; energies 
are given in units of GeV), 
for $d/r_o>1$ and (a) $(\epsilon L_s)_{45}/r_{16}=1$, 
(b) $(\epsilon L_s)_{45}/r_{16}=10^{-1}$, (c) ,
$(\epsilon L_s)_{45}/r_{16}=10^{-2}$ 
and (d) $(\epsilon L_s)_{45}/r_{16}=10^{-3}$. } 
\end{figure}

\begin{figure}
\vspace{10cm}  
\includegraphics{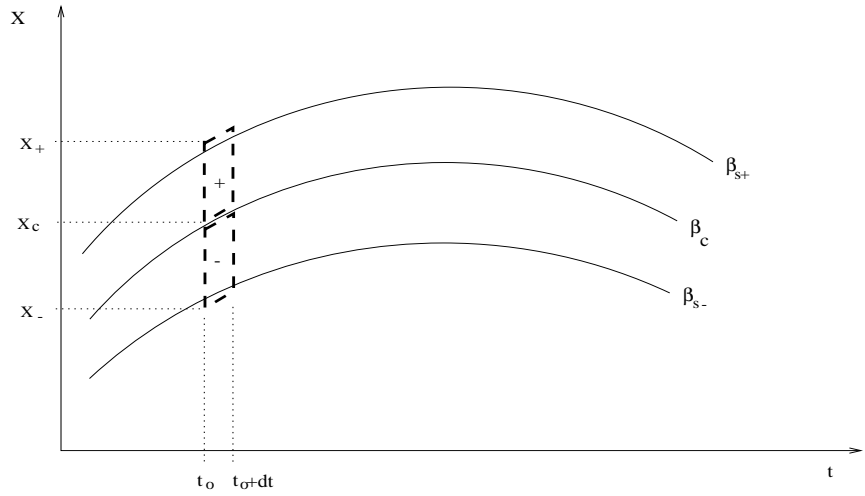}
\caption{Schematic diagram showing the world-lines of 
the forward shock ($\beta_{+}$), reverse shock 
($\beta_{-}$), and the contact discontinuity ($\beta_{c}$), and 
the corresponding contours of integration (dashed lines).  The 
region marked by the minus (plus) 
sign corresponds to the portion of the front leftward (rightward) 
of the contact discontinuity surface.}
\end{figure}
											
\end{document}